\newif\ifarxiv
\arxivtrue % or
\ifarxiv
	\documentclass[11pt]{article}
	\usepackage[total={6.5in, 9in}]{geometry}
	\usepackage{tgtermes}
	\usepackage[slantedGreek]{mathpazo}
	\usepackage[T1]{fontenc}
	\usepackage{amsmath}
	\usepackage[colorinlistoftodos]{todonotes}
\else
\fi

\DeclareMathOperator*{\mini}{minimize}
\DeclareMathOperator*{\argmax}{arg\,max}
\DeclareMathOperator*{\argmin}{arg\,min}

\usepackage{graphicx,float}%, float,wrapfig}
\usepackage{algorithm}% http://ctan.org/pkg/algorithms
\usepackage{algpseudocode}% http://ctan.org/pkg/algorithmicx
\usepackage[utf8x]{inputenc}

\graphicspath{{./fig/}{./fig/arxiv/}}

\begin{document}
\title{Practical Bayesian Optimization for Transportation Simulators}
\ifarxiv
	\author{
	Laura Schultz and Vadim Sokolov
	\footnote{\noindent Schultz is a PhD student at George Mason University, email: lschult2@gmu.edu. Sokolov is an assistant professor at George Mason University, email: vsokolov@gmu.edu}
	\\
	\textit{George Mason University}
	}
	\date{First Draft: December 2017\\This Draft: December 2018}
	\maketitle
\else
\fi

\begin{abstract}
We provide a method to 	solve optimization problem when objective function is a complex stochastic simulator of an urban transportation system. To reach this goal, a Bayesian optimization framework is introduced. We show how the choice of prior and inference algorithm effect the outcome of our optimization procedure. We develop dimensionality reduction techniques that allow for our optimization techniques to be applicable for real-life problems. We develop a distributed, Gaussian Process Bayesian regression and active learning models that allow parallel execution of our algorithms and enable usage of high performance computing. We present a fully Bayesian approach that is more sample efficient and reduces computational budget. Our framework is supported by theoretical analysis and an empirical study. We demonstrate our framework on the problem of calibrating a multi-modal transportation network of city of Bloomington, Illinois. Finally, we discuss directions for further research. 

%Simulators play a major role in analyzing multi-modal transportation networks. As complexity of simulators increases, development of calibration procedures is becoming an increasingly challenging task. Current calibration procedures often rely on heuristics, rules of thumb and sometimes on brute-force search. In this paper we consider an automated framework for calibration that relies on Bayesian optimization. Bayesian optimization treats the simulator as a sample from a Gaussian process (GP). Tractability and sample efficiency of Gaussian processes enable computationally efficient algorithms for calibration problems.  

\end{abstract}
	
\section{Introduction}\label{Introduction}
Suppose we have a transportation simulator $\psi(x,\theta)$ with $x$ being observable parameters, such as transposition network control strategies, day of week and weather and $\theta$ being unobserved latent variables, that represent, traveler's behavior preferences and transportation network attributes. Simulator outputs traffic counts (passengers, vehicles, pedestrians) at different locations on the network. Suppose we observe vector $y$
\begin{equation}
y(x_i)=\psi \left(x_i,\theta_i \right)+\epsilon \left(x_i\right)+e_i
\end{equation}
for different values of observable parameter $x$. Here, $e_i$ is the observation error and  inherent variations in the observed process; in practice, the observation error and residual variation cannot be separated out, and the computer model's inadequacy, $\epsilon$. In this paper we develop Bayesain techniques for solving an optimization problem 
\begin{equation}
\theta^* \in \argmin_{\theta \in A} L(\theta, D)
\end{equation}
Here $D = \{x_i,y_i\}_{i=1}^n$ is the set of observed input-output pairs and $L$ is a scalar function. The  constraint set $A$ encodes our prior knowledge about the feasible ranges of the model's parameters--for example, a traveler's value of time must be positive.

Function $L$ depends on the problem we are trying to solve. For example, for the calibration problem, the problem of adjusting model parameters $\theta$ so that simulated results match observed as close as possible, we have
\begin{equation}
L(\theta, D) =  \frac{1}{n}\sum_{i=1}^{n}||y_i - \psi(x_i,\theta)||_2^2
\end{equation}	
Which  is the divergence measure that quantifies the inadequacy between the observed data and the simulator's output. Since we do not know function $L$ we treat it as a black-box, which assume to only know the inputs and outputs of a process, can be leveraged.
\begin{figure}[H]
	\centerline{\includegraphics[width=0.45\textwidth]{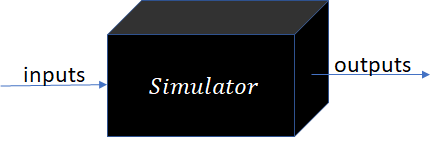}}
	\caption{Pictoral Representation of Black-box Treatment}
	\label{fig:BlackBoxOp}
\end{figure}

Unlike other similar approaches, we propose a fully Bayesian inference which allows us to estimate sensitivity to prior misspecification. Further, we develop linear and nonlinear dimensionality reduction techniques that allow us to solve practical large scale problems.  This paper's goal aside from introducing the dimensionality reduction techniques and Bayesian optimization algorithms, is to propose a framework under which our model can compared with other similar approaches.

We consider two major issues: choice of Bayesian priors and curse of dimensionality. To address the first issue, we develop a fully Bayesian approach in which parameters of the priors (hype-parameters) are inferred from the data during the optimization process. To address the second issue, we develop active subspace linear methods and deep learning nonlinear methods to find projection of parameter $\theta$ into lower-dimensional latent space in which the optimization problem is solved.

\section{Connection with Existing Literature}
Mobility dynamics in urban transportation systems are governed by a large number of travelers that act according to their utilities, preferences and biases. The resulting traveler behavior and observed mobility patterns cannot be adequately captured by single, closed-form models and, in practice, has resulted in the development and maintenance of complex, often stochastic computer simulations for dynamic insights into usage behavior and forecaster impacts of regional development.

With significant increases in data collection capabilities and utilization of greater computing power, transportation simulators are quickly growing in complexity as they attempt to more accurately approximate the processes they represent. These highly non-linear functions are becoming more sensitive to input variations and ballooning to unwieldy high dimensions without clear setting defaults. Furthermore, as higher quality data is integrated, adjustments to ensure previously developed simulators match observed field data becomes more complicated as knowledgeable developers transition to other projects and resource constraints limit the number of expensive setting experiments possible.

In response, we develop a Bayesian optimization framework~\cite{mockus1978application} for performing setting calibration of urban transportation system simulators. By finding a set of input parameters that lead to simulation outputs which match observed data, such as traffic counts, point-to-point travel times, or transit ridership, the simulators can be maintained, improved, and used confidently.

Traditional static assignment techniques~\cite{merchant_model_1978,merchant_optimality_1978}, such as the four-step model, allowed for a mathematically rigorous framework and set of efficient algorithms for transportation model calibration. However, more behaviorally realistic models that integrate dynamic traffic assignment, transit simulations, and dis-aggregate travel behavior paradigms do not allow for generalized universal solutions~\cite{peeta_foundations_2001}. Furthermore, modern simulators typically integrate multiple modules, such as dynamic traffic assignment, transit simulator, and activity-based models (ABM) for travel behavior, which are developed in ``isolation''. For example, discrete choice models are estimated using travel survey data before being implemented into the integrated ABM model. 

The calibration processes for these post-integrated models are, in practice, ad-hoc; however,
Numerous approaches for the calibration of simulation-based traffic flow models have been produced by treating the problem as an optimization issue\cite{garcia_decentralized_2007,zhang_lei_integrating_2013}. The lack of gradients in these models has led to mainly meta-heuristic methods being used, such as Simulation Optimization (SO)~\cite{chong_computationally_2017,zhang_efficient_2017}, Genetic Algorithm (GA)~\cite{cheu_calibration_1998,ma_genetic_2002}, Simultaneous Perturbation Stochastic Approximation (SPSA)~\cite{lu_enhanced_2015,cipriani_gradient_2011,lee_new_2009}, and exhaustive evaluation~\cite{hale_optimization-based_2015}, with relative success~\cite{yu_calibration_2017}.

Alternatively, Bayesian inference methods provide a confidence value and analytic capability not necessarily produced by other general-purpose approaches for non-parametric linear and non-linear modeling; although limited, their application in transportation has been successful \cite{hazelton_statistical_2008,flotterod_general_2010,flotterod_bayesian_2011,zhu_calibrating_2018}. One of the first statistical methodologies to address the analysis of computer experiments in general can be found in \cite{sacks_design_1989}, which introduces a kind of stochastic process known as a Gaussian Process(GP) for use with Bayesian inference\cite{gramacy_adaptive_2009,snoek_input_2014,danielski_gaussian_2013,rasmussen_gaussian_2006,romero_navigating_2013}. The application of GP regression towards calibration was pioneered by \cite{kennedy_bayesian_2001}, where the concept was deployed as surrogate model, or emulator, which estimated the sources of uncertainty between the simulation and true process to improve the prediction accuracy for unverified variable settings. \cite{higdon_computer_2008} further expands this framework to address high-dimensional problems with high-dimensional outputs by combining it with Markov chain Monte Carlo(MCMC) methods; others have begun integrating these with Machine Learning techniques\cite{snoek_practical_2012,rasmussen_gaussian_2006}.

However, the primary focus for several of these applications centered on making the most accurate predictions rather than on aligning the simulators themselves. Successful calibration will require a balance the exploration of unknown portions of the input sample space with the exploitation of all known information. \cite{chaloner_bayesian_1995} and \cite{ryan_contributions_2014} discuss extensively several Bayesian utility functions and their non-Bayesian Design of Experiments (DOE) equivalents.

In addition, addressing the exponential increase in the dimensions of transportation simulators is becoming more paramount as models become more detailed and environments grow larger. The dis-aggregation of an origin-destination matrix into individual trip-makers through activity-based decision modeling (ABDM) is one approach to decreasing the complexity of traffic flow dynamics in complex network infrastructures~\cite{nagel_agent-based_2012}. Pre-processing methods such as Principal Component Analysis (PCA)~\cite{djukic_efficient_2012} have been used in transportation to further reduce dimensionality with minimal cost to accuracy. 

%We build on these existing works in simulation-based optimization literature both in transportation and other engineering fields by formalizing a Bayesian optimization framework to calibrate complex transportation simulators. Our contributions are:
%\begin{itemize}
%	\item Formulation of Gaussian Process Bayesian methods for transportation optimization
%	\item Experiment Design algorithms for active conservation of restricted resource allocations
%	\item Dimensionality reduction techniques for complexity control
%	\item Integration of Deep Learning with Gaussian Process for improved mean function
%\end{itemize}

%The remainder of this paper is organized as follows: Section 2 defines a notation set that will be used throughout the document and details the framework and approaches to calibration; Section 3 addresses the need for dimensionality reduction; Section 4 outlines the contribution enhancement to the Gaussian Process structure; Section 5 provides an illustrative example; and Section 6 offers conclusions and avenues for further research.

\section{Bayesian Optimization} 

\begin{figure}[H]
	\centerline{\includegraphics[width=0.8\linewidth]{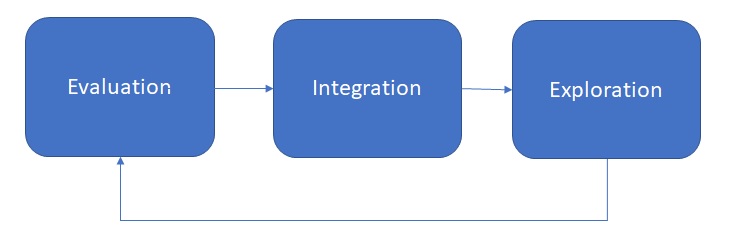}}
	\caption{Three Stages of Calibration Framework}\vspace*{-6pt}
	\label{fig:calibrationframework}
\end{figure}

We take a Bayesian approach to the optimization problem which, includes
\begin{enumerate}
	\item Put prior on continuous functions $C[0,\infty]$
	\item Repeat for $k=1,2,\dots$
	\item Evaluate $\psi$ at $\theta^{k}_1,\ldots,\theta^{k}_{n_k}$
	\item Compute a posterior (integrate)
	\item Decide on the next batch of points to be explored $\theta^{k+1}_1,\ldots,\theta^{k+1}_{n_{k+1}}$
\end{enumerate}
Our framework is designed to execute Bayesian optimization algorithms in distributed computing environments and can be broken down, as shown in Figure \ref{fig:calibrationframework}, into three reiterative stages: Evaluation, Integration, and Exploration.

For the posterior calculation, we use the Bayes rule and incorporate  the evidential results collected in the Evaluation stages, which states that the posterior probability of a model, given a set of evidential data, is proportional to the likelihood of the evidence given the model multiplied by the prior probability of the model:
\begin{equation}
P(\mathrm{Model} \mid \mathrm{Evidence}) \propto P(\mathrm{Evidence} \mid \mathrm{Model})P(\mathrm{Model})
\end{equation}
For a more in-depth review of Bayesian optimization and previous work in black-box methodologies, refer to \cite{brochu_tutorial_2010}. 

We use Gaussian Process prior which is a non-parametric model that defines distribution over continuous and is functions and  is fully characterized by its mean function and its positive-definite kernel, or covariance function~\cite{gramacy_particle_2011}.  Formally, $g(x)$ is a GP if, for any finite-dimension $d$ subset of its random variables, the joint distribution $g_{x_1,..,x_d}$ produces a Gaussian Distribution.

It may help to consider a single function $g(x)$ drawn from one of these GPs as an infinitely-long vector drawn from an extremely high-dimensional Gaussian~\cite{rasmussen_gaussian_2006}. In application, having a GP described in the same manner as a Gaussian Distribution, by an infinite-dimensional scalar mean vector ($\infty \times 1$) and infinite-squared covariance matrix ($\infty \times \infty $), would be impractical. A GP over an infinite collection of random variables, is, instead, described with a mean function $m$ and a covariance function $k$ 

\begin{equation}
g\left(x\right) \sim \mathcal{GP}(m(x),k(x,x^{\prime}))
\end{equation}

where
\begin{equation}
m(x) = E[x], \qquad k(x,x^\prime) = E[(x-m(x))(x^\prime - m(x^\prime)]
\end{equation}

The covariance function $k(x,x^\prime)$ of a distribution specifies the spatial relationship between two input sets $x$ and $x^\prime$; that is, it acts as an information source detailing the degree at which a change in the distribution value at $x$ will correlate with a change in the distribution value at $x^\prime$. The mean function can be represented as any function; however, the covariance function must result in a positive semi-definite matrix for any of its subset Gaussian Distributions. 

This positive semi-definite requirement for covariance matrices is identical to Mercer's condition for kernels and, therefore, any kernel-based function is a valid covariance. The most commonly applied kernel is known as a Squared Exponential or Radial Basis kernel, which is both stationary and isotropic and results in a homogeneous and smooth function estimate\footnote{smooth covariance structures encode that the influence of a point on its neighbors is strong but nearly zero for points further away}.
\begin{list}{}{}
	\item
	\begin{equation}
	k_{\mathrm{SE}}(x,x^\prime \mid \omega=[\sigma,\lambda]) = \sigma^2\exp{\left[-\frac{1}{2}{\left(\frac{x-x^{\prime}}{\lambda }\right)}^{2}\right]}
	\end{equation}
	
	where $\sigma$ is the output variance and $\lambda$ is the length scale, the kernel hyperparameters represented collectively by the symbol $\omega$.
	
\end{list} 
A second common choice is known as the Mat\'ern covariance function, which provides less smoothness through the reduction of the covariance differentiability.

\begin{list}{}{}
	\item 
	\begin{equation}
	k_{\mathring{Matern}}(x,x_{\prime} \mid \omega=[\sigma,\lambda,v]) = \sigma^2\frac{{2}^{1-v}}{\Gamma(v)}\left[\frac{\sqrt{2v}|x-x^{\prime}|}{\lambda}\right]^v K_v\left[\frac{\sqrt{2v}|x-x^\prime|}{\lambda}\right]
	\end{equation}
	
	where $\sigma^2$ is the output variance, $\Gamma$ is the gamma function, $K_{v}$ is a modified Bessel function, and $\lambda$ and $v$, non-negative hyperparameters for length scale and smoothness respectively.
\end{list} 

Other common but less used forms include periodic, which can capture repeated structure, and linear, a special case of which is known to model white noise. General references for families of correlation functions and possible combinations are provided by \cite{abrahamsen_review_1997} and \cite{duvenaud_automatic_2014}.

We account for process variance and measurement errors by adding to the covariance function \cite{gramacy_cases_2012}:
\begin{equation}
L(\theta,D)\sim \mathcal{GP}(m(\theta),k(\theta,{\theta}^{\prime}) + \sigma_e^2\delta_{\theta'} \mid \Omega=[\omega,\sigma_e])
\end{equation}
where $\delta_{\theta'}$ is the Kronecker delta which is one if $\theta = \theta^{\prime}$ and zero otherwise; $\sigma_e^2$ is the variation of the error term; and $\omega$ encapsulates the chosen kernel's hyperparameters.

This formulation continues to result in the variance of an input set $\theta_j$ increasing away from the nearest alternative input $\theta_i$ as before; however, it no longer results in zero if $\theta_j = \theta_i$ but rather $\delta\sigma_e^2$.

\begin{figure}
	\centering
	\begin{tabular}{cc}
		\includegraphics[width=0.3\linewidth]{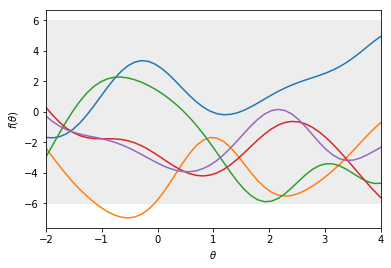} &
		\includegraphics[width=0.3\linewidth]{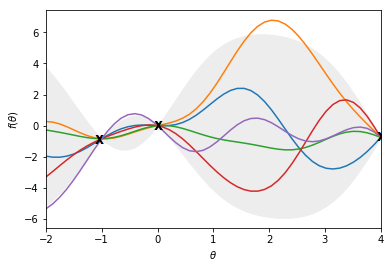}\\
		(a) Draw from prior & (b) Draw from posterior
	\end{tabular}
	\caption{(a) shows five possible functions from a Gaussian Process prior using the Squared Exponential Kernel. (b) shows five possible functions drawn from the posterior distribution resulting from the prior being conditioned on several observations indicated by an 'x' marker. In both plots, the shaded area represents the 95\% confidence interval.}
	\label{fig:BayesGP}
\end{figure}

To sample from this prior distribution, function values $f_t$ would be drawn for $t$ input sets ${\theta_{1:t}}$ according to a Gaussian Distribution $\mathcal{N}(\mu = m(\theta), K= k(\theta,\theta^\prime)+\sigma_e^2\delta_{\theta'})$ with the kernel matrix given by:
\begin{equation}
K = \left[\begin{array}{ccc}k(\theta_1,\theta_1)+\delta\sigma^2 & \dots & k(\theta_1,\theta_t)\\ \vdots & \ddots & \vdots \\k(\theta_t,\theta_1) & \dots & k(\theta_t,\theta_t)+\delta\sigma^2\end{array}\right]
\end{equation}

where $k(\cdot,\cdot)$ is the chosen kernel function. Figure \ref{fig:BayesGP}(a) shows an example of potential representative functions given only a prior knowledge of the simulator.

The posterior is conditioned from a joint distribution between the input sets which have been evaluated, designated as $\theta_{ev}$, and the sets which have not, designated as $\theta_\ast$:

\begin{equation}
p\left(\left[\begin{array}{c}L(\theta_{\mathrm{ev}},D))\\ L(\theta_*,D))\end{array}\right]\right) \sim \mathcal{N}\left(\mu = \left[\begin{array}{c}
\mu_{ev} \\ \mu_{ev}\ast\end{array}\right], K = \left[\begin{array}{cc}
K_{ev,ev} \ K_{ev,\ast} \\ K_{\ast,ev} \ K_{\ast,\ast}\end{array}\right]\right)
\end{equation}

To sample from this posterior distribution, function values, $L(y,\psi(\theta_{t}))$, would be computed for $t$ input sets, ${\theta_{1:t}}$, according to a conditional Gaussian Distribution 

\begin{equation}
p[L(\theta_*,D))  \mid  \theta_\ast, \ \theta_{ev}, \ L(\theta_{\mathrm{ev}},D)), \ \Omega] \sim \mathcal{N}(\mu_{post},K_{post})
\end{equation} 

with the following summary statistics:

\begin{equation}
\mu_{post} = \mu_\ast + K_{\ast,ev}K^{-1}_{ev,ev}[L(\theta_{\mathrm{ev}},D))-\mu_\ast], \qquad K_{post} = K_{\ast,\ast} - \ K_{\ast,ev}K^{-1}_{ev,ev}K_{ev,\ast}
\end{equation}

Figure \ref{fig:BayesGP}(b) shows an example of the potential representative functions given a few data points evaluated by the simulator.

\subsection{Exploration}
The \textbf{exploration} stage provides a recommended input set for the next iteration. Typical Design of Experiment (DOE) methods such as randomization, factorial, and space-filling designs begin with the entire state space and provide a pre-determined list of candidates that should be run regardless of the previous evaluation outcomes. However, computational cost of running transportation simulators require attention be given to minimizing the number of samples without compromising the final recommendation and, consequently, candidates need to be ordered in such a manner that redundant sampling in areas of the state space already adequately mapped does not occur. 

A machine learning technique known as \textit{active learning}, also known as \textit{optimal experimental design} in DOE, provides such a scheme. A utility function (a.k.a acquisition function) is built to balance the exploration of unknown portions of the input sample space with the exploitation of all information gathered by the previous stages and cycles, resulting in a prioritized ordering reflecting the motivation and objectives behind the calibration effort. Formally, this is written as \cite{ryan_contributions_2014}:
\begin{equation}
\theta^+=\argmax_{\theta \in D} ~ E[U\left(\theta,L,\Omega)\right]
\end{equation}
where $\theta^+$ is the optimal design choice, or input set, decision from the potential candidate set $D$; $U$ is the chosen utility function reflecting the experiment's purpose (inference or prediction of unknown parameters $\theta$); $L(\theta)$ is the surrogates objective function; and $\Omega$ represents the accompanying hyperparameters of the surrogate model.	

The expectation of the utility function is taken over the posterior distribution calculated in the previous Integration stage and ,thus, the optimal design choice will be the yet-evaluated input set which maximizes the posterior expected utility. It should be noted that this framework does not aim to specify a single utility function to be used in all employed circumstances but to provide context behind which active learning utilities should be used for specific calibration situations.  

\subsubsection{Acquisition Functions}
Within the active learning framework, utility functions are often referred to as acquisition functions. With several different parameterized acquisition functions in the literature, it is often unclear which one to use. Some rely solely on exploration-based objectives, choosing samples in areas where variance is still large, or exploitation-based objectives, choosing samples where the mean is low, while others lie somewhere in between. 

Determining the best acquisition function depends upon the overall purpose of the experiments. For example, predicting future values rely on minimizing variance across the state space and result in the chosen acquisition function skewing towards exploration while parameter estimation concentrates on finding the lowest or highest mean of the function through a bias towards exploitation. This framework concentrates on the later.

\cite{chaloner_bayesian_1995} discusses extensively several Bayesian utility functions and their non-Bayesian DOE equivalents. Below are a few more widely used acquisition functions for Bayesian designs:
\begin{figure}
	\centering
	\includegraphics[scale=0.5]{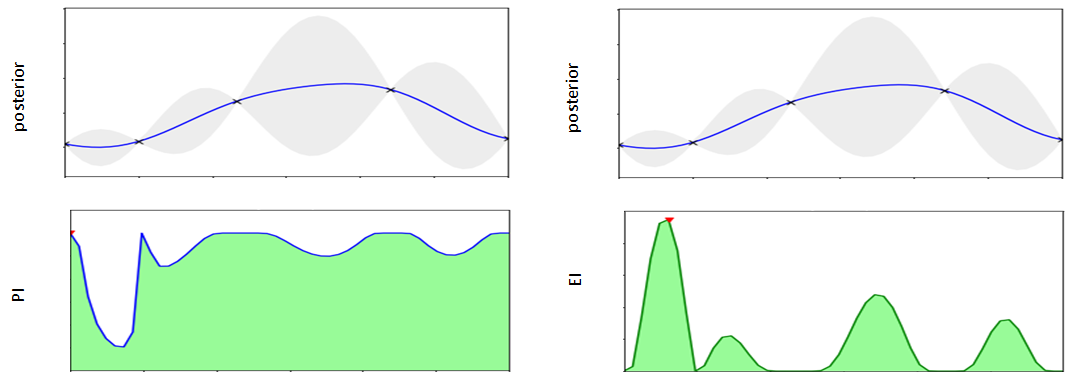}
	\caption{ Examples of acquisition functions. The GP posterior is displayed at the top. The underneath plots display the Probability of Improvement (PI) and Expected Improvement (EI) acquisition functions for the GP on the left and right, respectively. The maximum of each function is marked by a red triangle.}
	\label{fig:AcqEx}
\end{figure}

\paragraph{Probability of Improvement}
The Probability of Improvement(PI) function was first proposed by Harold Kushner \cite{kushner_new_1964} and proposes a simple 0-1 utility:

\begin{equation}
U(\theta^\ast) = 
\begin{cases}
0, \ L(\theta^\ast) > L(\theta^+) \\
1, \ L(\theta^\ast) \le L(\theta^+)
\end{cases}
\end{equation}

where $\theta^\ast$ is a potential candidate from the unevaluated set of possible samples and $\theta^+ = \arg \max_{\theta \in \theta_{eval}} L(\theta_i)$, the best solution from the evaluated set which produces the current minimum value found.

When the expectation is taken over the predictive distribution, the point with the highest probability of providing a smaller value becomes the recommended candidate. For application to our Gaussian posterior distribution, this can be written as:

\begin{equation}
\begin{split}
 PI(\theta^\ast) = & E[U(\theta^\ast)] \\
 = & \int_{-\infty}^{L(\theta^+)} \mathcal{N}\left(L;\mu_{post},K_{post}\right) dL\\
 = & P(L(\theta^\ast) \le L(\theta^+))  \\
 = & \Phi \left(\frac{\mu_{post}(\theta^\ast) - L(\theta^+)}{K_{post}(\theta^\ast)}\right)
\end{split}
\end{equation}

where $\Phi(\cdot)$ is the Gaussian cumulative distribution function and $\mathcal{N}\left(L;\mu_{post},K_{post}\right)$ is the predictive posterior distribution outlined in Section \ref{Noisy}.

In the original form above, this function is purely exploitative and relies heavily on the initial placement of the original samples. A simple modification exists, however, to encourage exploitation by adding a trade-off hyperparameter, $\lambda \ge 0$, to the left-hand side.
\begin{equation}
 PI(\theta^\ast) = P(L(\theta^\ast) \le L(\theta^+ - \lambda)) 
\end{equation}
Although it can be set at the user's discretion, Kusher recommended a decaying function to encourage exploration at the beginning and exploitation thereafter. 

\paragraph{Expected Improvement}
Defined by Mockus \cite{mockus_application_1978}, this utility function places emphasis on the size of the improvement to avoid getting stuck in local optimas and under-exploration. 

\begin{equation}
U(\theta^\ast) = \max [0, L(\theta^+) - L(\theta^\ast)]
\end{equation}

where $\theta^\ast$ is a potential candidate from the unevaluated set of possible samples and $\theta^+ = \arg \max_{\theta \in \theta_{eval}} L(\theta_i)$, the best solution from the evaluated set which produces the current minimum value found.

When the expectation is taken over the predictive distribution, the point now with the highest probability of providing the greatest amount of improvement becomes the recommended candidate. For application to our Gaussian posterior distribution, this can be written as:

\begin{equation}
\begin{split}
EI(\theta^\ast) = & E[U(\theta^\ast)] \\
= & \int_{-\infty}^{(L(\theta^+)-L(\theta^\ast)L(\theta^+)} \mathcal{N}\left(L;\mu_{post},K_{post}\right) dL\\
= & (L(\theta^+)-L(\theta^\ast)\Phi \left(\frac{\mu_{post}(\theta^\ast) - L(\theta^+)}{K_{post}(\theta^\ast)}\right) + K_{post}(\theta^\ast)\phi \left(\frac{\mu_{post}(\theta^\ast) - L(\theta^+)}{K_{post}(\theta^\ast)}\right) \\
= & K_{post}(\theta^\ast) [u\Phi(u) + \phi(u)]
\end{split}
\end{equation}

where $u = \frac{L(\theta^+)-\mu_{post}(\theta^\ast)}{K_{post}(\theta^\ast)}$, $\Phi(\cdot)$ is the normal cumulative distribution, $\phi(\cdot)$ is the normal density function, and $\mathcal{N}\left(L;\mu_{post},K_{post}\right)$ is the predictive posterior distribution outlined in Section \ref{Noisy}.

With this formulation, a trade-off of exploitation and exploration is automatic; the expected improvement can be influenced by a reduction in the mean function (exploitation) or by increasing the variance (exploration).

\paragraph{Upper Confidence Bound}
A recently developed but useful acquisition function, the Upper Confidence Bound (UCB) method moves away from the expected evaluation of a utility function. Instead, it relies on the theoretical proofs that, under specific conditions, the iterative application will converge to the global minimum of the interested function:

\begin{equation}
UCB(\theta^\ast) = \mu_{post}(\theta^\ast) - \beta \sqrt{K_{post}(\theta^\ast)}
\end{equation}

where $\beta > 0$ is a trade-off hyperparameter and $[\mu_{post},K_{post}]$ are the summary statistics of the predictive posterior distribution outlined in Section \ref{Noisy}.

\section{Dimensionality Reduction}
\label{Dimensionality Reduction}

The contribution of a single sample to the understanding of evaluation space decreases as the space between samples becomes larger and dimensions grow. As a result, the cost of constructing a surrogate function for a simulator increases exponentially as the dimension of the input space, or input parameters, increases; this is often referred to as the 'curse of dimensionality'. In practice, however, high dimensional data possesses a natural structure within it that can be successfully expressed in lower dimensions. Known as dimension reduction, the effective number of parameters in the model reduces and enables successful analysis from smaller data sets. 

In this paper, two Latent Variable Model (LVM) techniques are explored: Active subspaces, a linear method, to identify and segregate the input dimensions into important, or active, and less-important, or inactive, directional categories.\cite{constantine_active_2015}; Combinatorial deep learning architecture, a non-linear method, to compose a nested structure of univariate functions to approximate the input-output relations.

By identifying a reduced dimensional space, analysis methods such as Gaussian surrogate techniques, become more powerful and favorable.

\subsection{Active Subspaces}

Active subspace identifies linear combinations of inputs which significantly influence, on average, the output of the simulation when minor adjustments are made.  These combinations are derived from the gradients between the inputs and quantities of interest decomposed into eigenvector principal components, $W$:
\[
C = \int (\nabla_{\theta}L)(\nabla_{\theta}L)^Td\theta = W\Lambda W^T
\]
where $C$ is a sum of semi-positive definite rank-one matrices, $W$ is the matrix of eigenvectors, and $\Lambda$ is the diagonal matrix of eigenvalues in decreasing order.

This transformation is defined in such a way that the first principal component accounts for as much of the variability in the data as possible, and each succeeding component in turn has the highest variance possible under the constraint that it is orthogonal to the preceding components.

\begin{figure}[]
	\begin{tabular}{cc}
		\hspace*{20mm}\includegraphics[width=0.35\linewidth]{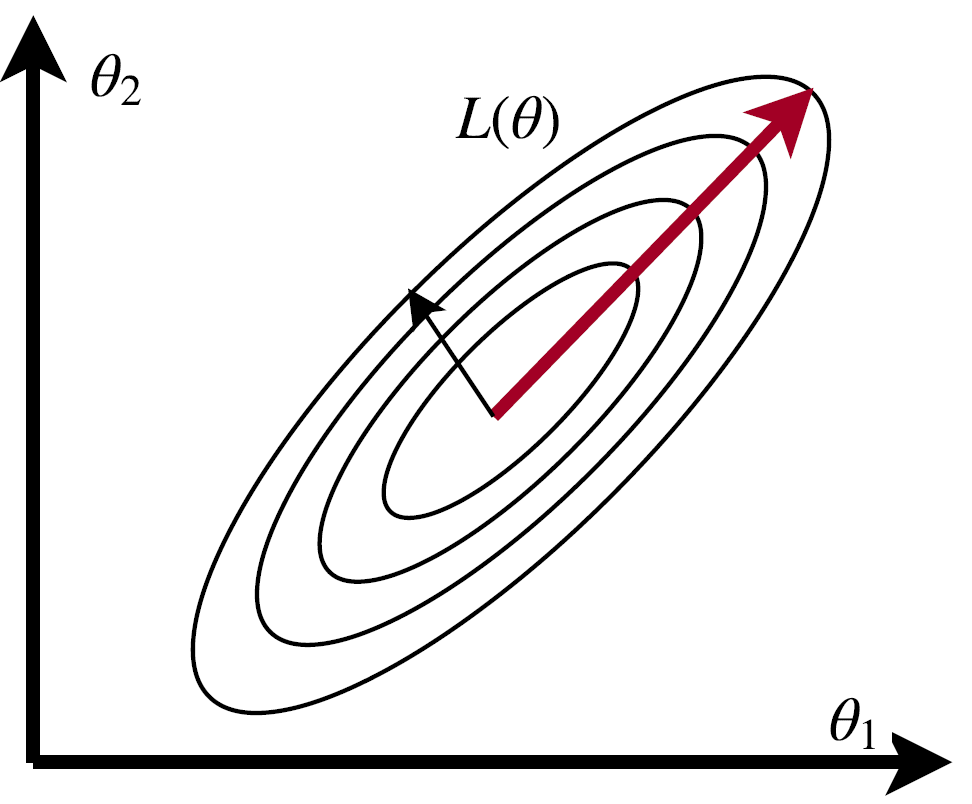} &
		\hspace*{10mm}\includegraphics[width=0.3\linewidth]{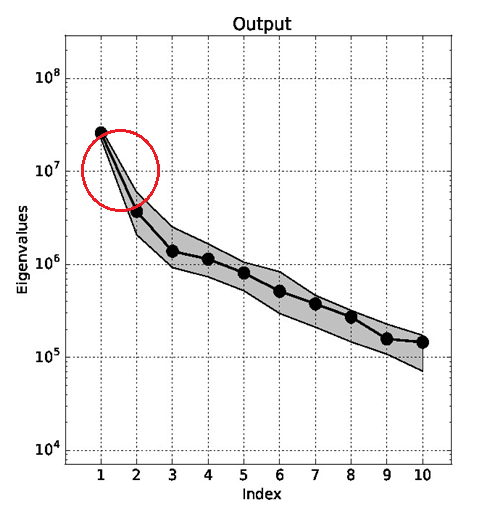} \\
		(a) Principal Components   & (b) Active Subspace Detection  \\	
	\end{tabular}
	\caption{(a) A 2-dimensional example of Principal Components. The magnitude of the red arrow indicates that most of the variability can be explained by the direction and is, therefore, the first principal component. (b) An example of the eigenvalues plotted for active subspace detection. The largest eigenvalue gap is designated by a red circle. The eigenvalues to the left compose the active subspace.}
	\label{fig:activesubspaces}
\end{figure}

To visualize, these principal components can be imagined as an $m$-dimensional ellipsoid. If any axis of the ellipsoid is small, then the variance along that axis, or direction, is also small, and by omitting that axis and its corresponding principal component from our representation of the data set, we lose only a commensurately small amount of information.  An example is shown in Figure \ref{fig:activesubspaces}(a).

An observant reader may note this concept is strikingly similar to the Principal Component Analysis (PCA) method. Indeed, the primary difference between the two centers around the criteria used to determine what eigenvalues are significant. PCA will choose the eigenvalues which, when summed, reach a pre-specified proportion of all eigenvalues. Active Subspace plots the components on a log-scale and a dramatic change in the eigenvalue space, documented as a \textit{gap}, is looked for; see Figure \ref{fig:activesubspaces}(b) for a demonstration.  To the left are the active subspaces and to the right are the inactive subspaces.

For the calibration framework outlined in this paper, the chosen active subspace is applied to the surrogate model, $L(y,\psi(\theta))$, outlined in Section \ref{Integration} as follows:
\begin{equation}
L\left(y,\psi(\theta)\right)=L\left(\hat{W}_{1}v+\hat{W}_{2}z\right)
\end{equation}

where $\hat{W}_{1}$ contains the first $n$ eigenvectors found to be active, $\hat{W}_{2}$ contains the remaining $m-n$ eigenvectors, $v$ contains the active subspace variables, and ${z}$ contains the inactive subspace variables.

Consider the following algorithmic procedure to determine the active subspace for a generic function $f(x)$:
\begin{enumerate}
	\item Normalize the input parameters around zero to remove units and prevent larger inputs from biasing the subspace
	\item Draw $i= 1..N$ independent samples according to a per-determined sampling method $\rho(x)$, such as a uniform sampling over a hypercube
	\item For each sample $x_i$, compute the quantity of interest $f_i = f(x_i)$
	\item Use a gradient approximation method to determine the gradients $\nabla_x f_i$. 
	For example, use least-squares to fit the coefficients $\hat{\beta_0}$ and $\hat{\beta}$ of a local linear regression model
	\begin{center}
		\begin{equation}
		{f}_i\approx {\hat{\beta}}_{0}+\hat{\beta}^{T}x_i
		\end{equation}
		\begin{equation}
		\nabla_x f_i = \hat{\beta} 
		\end{equation}
	\end{center} 
	\item Compute the matrix $\hat{C}$ and its eigenvalue decomposition
	\begin{center}
		\begin{equation}
		\hat{C}=\frac{1}{M}\sum_{i=1}^{M} (\nabla_x f_i)(\nabla_x f_i)^{T} = \hat{W} \hat{\lambda} \hat{W}^{T}
		\end{equation}
	\end{center}
	
	Where $\hat{W}$ is the matrix of eigenvectors, and $\hat{\lambda}$ = $diag(\hat{\lambda}_{1},...,\hat{\lambda}_{m})$ is the diagonal matrix of eigenvalues ordered in decreasing order.
\end{enumerate}

For more information behind the derivation of this procedure and its variations, see \cite{constantine_active_2015}.

Once the procedure above is completed, the eigenvalues should be plotted and large gaps identified, as shown in Figure \ref{fig:activesubspaces}(b). Because the error is inversely proportional to the corresponding gap in the eigenvalues, the dimension $n$ should be chosen such that the largest eigenvalue gap is the separator. These $n$ directions, where $n$ is less then the original dimension set $m$, are considered the active subspace; the remaining $m-n$ directions are the inactive subspace. 

If no gap can be found, compiling larger sets of eigenvalues or sampling more within the current eigenvalue framework to increase the eigenvalue accuracy is suggested.

For example, if the gap between the 2nd and 3rd eigenvalues is larger than the gap between the first and second eigenvalues, the estimates of a 2-dimensional active subspace is more accurate than a 1-dimensional subspace, as shown in Figure \ref{fig:activesubspaces}(b). 

Using this formulation, the inactive variables($z$) can be fixed at nominal values while the important variables are varied. This follows the concept that small perturbations in $z$ change $f$ relatively little, on average. For situations in which the inactive eigenvalues are exactly zero, $f(x)$ can be projected onto the active subspace without regard to the inactive subspace by setting $z=0$:

\begin{equation}
 f\left(\theta\right)\approx f(\hat{W}_{1}v) \approx f (\hat{W}_{1}\hat{W}_{1}^{T} \theta)
\end{equation}

However, for situations in which the inactive eigenvalues are small but not zero, the active subspace structure should be constructed by optimizing over $v$ such that:

\begin{equation}
\min_{v \in V} \left[\begin{array}{cc}
\mathrm{minimum}_{z \in Z} & f\left(\hat{W}_{1}v+\hat{W}_{2}z\right) \\ \mathrm{subject \ to} &  x_{lb}-\hat{W}_{1}v \le \hat{W}_{2}z \le x_{ub} - \hat{W}_{1}v
\end{array}\right]
\end{equation}

where $x_{lb}$ and $x_{ub}$ are the lower and upper bounds of the original $x$ state-space, respectively.

\subsection{Deep Neural Networks}

The relationship between a (high dimensional) set of inputs, $x= (x_1 , \ldots , x_p ) \in R^p $, and a multi-dimensional output, $y$ is often depicted through a multivariate function,$y=f(x)$. However, as the relationship between inputs and outputs become more complicated, either $f(x)$ must grow to accommodate or another method of depicting this relationship must be found.
 
A Deep Learning Neural Network is a pattern matching model which approaches this complication by replacing the single function depiction with an interconnected network of simpler functions. This paradigm finds its roots in the way the biological nervous systems, such as the brain, process information: a collection of units known as \textit{neurons} communicate with one another through synaptic connections. 

The receiving neuron, or post-synaptic neuron, processes the provided information through a simple function before sending it on; a group of neurons which perform the same simple function on the same set of inputs is known as a layer. By composing $L$ layers, a deep learning predictor, known as a Multi-layer Perceptron network, becomes
\begin{equation}
\begin{split}
\hat y = & F_{W,b}(x) = (f_{w_0,b_0}^0 \circ \ldots \circ _{w_L,b_L}^L)(x)\\
f^i_{w_i,b_i} = & f_i(w_ix_i +b_i) \quad \forall \quad i \in [0,L].
\end{split}
\end{equation}

Here $f_i$ is a univariate activation function such as $\tanh$, sigmoid, softmax, or Relu. Weights $w_l \in R^{p_l \times p_l'}$ and offsets $b_l \in R$ are selected by applying stochastic gradient descent (SGD)~\footnote{It should be noted that other optimization algorithms to determine the correct weights and biases do exist. Some build upon backpropagation like SGD, while others diverge to leverage  more sophisticated concepts. In this paper, however, backpropagation is sufficient} to solve a regularized least squares optimization problem given by

\begin{equation}
\mini_{W,b} \; \sum_{i=1}^N ||y_i - F_{W,b} ( x_i ) ||_2^2 + \phi ( W,b )
\end{equation}

Here $\phi $ is a regularization penalty on the network parameters (weights and offsets).

An auto-encoder is a deep learning routine which trains the architecture to approximate $x$ by itself (i.e., $x$ = $y$) via a bottleneck structure. This means we select a model $F_{W,b}(x)$ which aims to concentrate the information required to recreate $x$. Put differently, an auto-encoder creates a more cost effective representation of $x$. For example, under an $L_2$-loss function, we wish to solve 
\begin{equation}
\mini_{W,b} ||F_{W,b}(x)-x||_2^2
\end{equation}

subject to a regularization penalty on the weights and offsets. In an auto-encoder, for a training data set $\{x_1,\ldots,x_n\}$, we set the target values as $y_i = x_i$. A static auto-encoder with two linear layers, akin to a traditional factor model, can be written via a deep learner as
\begin{equation}
\begin{split}
a^{(1)} = &x\\
a^{(2)} = &f_1(w^{(2)}a^{(1)} + b^{(2)})\\
a^{(3)} = & F_{W,b}(x) = f_2(w^{(3)}a^{(2)} + b^{(3)}),
\end{split}
\end{equation}

where $a^{(i)}$ are activation vectors. The goal is to find the weights and biases so that size of $w^{(2)}$ is much smaller than size of $w^{(3)}$.

Within the calibration framework, two objectives must be realized by the neural network:

\begin{enumerate}
	\item A reduced dimension subspace which captures the relationship between the simulator inputs and outputs must be found and bounded in order for adequate exploration of the state-space to determine the next useful evaluation point
	\item Given that the recommended evaluation points are expressed in the reduced dimension sample space, the network must be able to recover a projected sample back to the original space to allow for simulator evaluation
\end{enumerate} 

\begin{figure}[]
	\centering
	\includegraphics[width=0.6\linewidth]{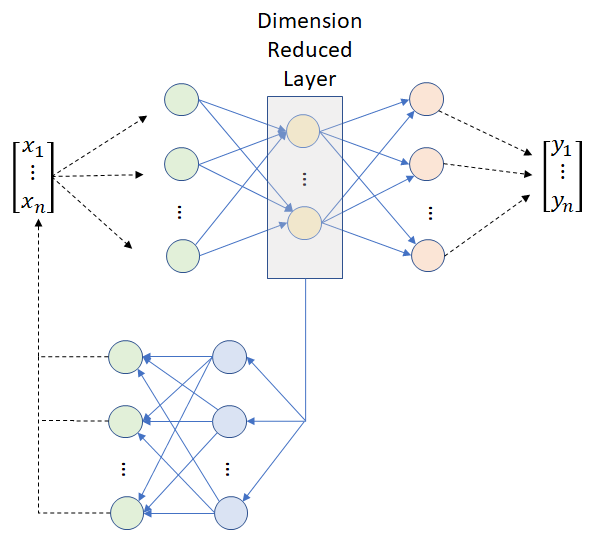}
	\caption{Graphical Representation of the Combinatorial Neural Network for Calibration}
	\label{fig:MLPAuto}
\end{figure}

To address these objectives, we combine an MLP architecture to capture a low-dimensional representation of the input-output with an autoencoder architecture to decode the new subspace back to the original. We will then be able to run the calibration framework's optimization algorithms inside the low dimensional representation of the input parameter space to address the curse of dimensionality issue and convert the recommended sample into a usable form for simulator evaluation. The Autoencoder and MLP share the same initial layers up to the reduced dimension layer, as shown in Figure \ref{fig:MLPAuto}.

The activation function used is a bounded $RELU$, which provides a bounded range for search purposes. Additionally, to quantify the discrepancies during training of the neural network, the following loss functions are used:
\begin{enumerate}
	\item The MLP portion of the architecture uses the mean squared error function $L$
	\begin{equation}
	 L(\theta,D)= \frac{1}{N}\sum_{i=1}^{N}(y - \psi(\theta_i))^2
	\end{equation}
	where $\psi(\theta)$ represents the predicted values produced by the neural network for the simulator's output $y$ given the input set $\theta$
	\item The Autoencoder portion of the architecture uses the mean squared error function $L$ and a quadratic penalty cost $P$ for producing predicted values outside of the original subspace bounds since the simulator cannot evaluate a point that does not lie in the plausible domain
	\begin{equation}
	 D\left[\psi(\theta)\right] = \max[0,\psi(\theta_i)-x_u]^2 + \max[0,x_l-\psi(\theta_i)]^2
	\end{equation}
	where $\psi(\theta)$ represents the predicted values produced by the neural network for the simulator's input $x$ given the input set $\theta$, $x_u$ represents the input set's upper bound, and $x_l$ represents the input set's lower bound
\end{enumerate}

The authors would like to note that, while it can be separated and trained sequentially, this architecture should be trained using a combined loss function, i.e. the loss from the errors in the predicted $y$ values added to the loss from the errors in the predicted $x$ reconstruction values. When trained in this manner, the network is encouraged to ensure a unique representation of the relationship is captured in the reduced dimension layer and results in significantly more successful recommended samples during the exploration stage. 

\section{Improved Mean Function}

Thus far, additional statistical approaches have only been utilized for the dimensionality reduction portion of the calibration framework. However, the structural assumption of a zero-mean function typical of GP applications allows for further leverage. Indeed, an educated mean function which can provide additional relational information to the distinct application of interest allows for a broadly useful approach. 

Recall from Section \ref{Integration} that the posterior, predictive distribution possesses the following summary statistics:

\begin{equation}
\mu_{post} = \mu_\ast + K_{\ast,ev}K^{-1}_{ev,ev}[L(\theta_{\mathrm{ev}},D))-\mu_\ast], \qquad K_{post} = K_{\ast,\ast} - \ K_{\ast,ev}K^{-1}_{ev,ev}K_{ev,\ast}
\end{equation}

When non-zero, the surrogate function's mean,$\mu_\ast$, not only influences the predictive mean, but also the predictive covariance. In addition, the acquisition functions also rely heavily on the predicted mean of the potential candidate; for example, the Expected Improvement utility for any Guassian is written as
\begin{equation}
EI(\theta) = \sigma(\theta) [u\Phi(u) + \phi(u)]
\end{equation}
where $u = \frac{f(\theta^+)-\mu}{\sigma(\theta)}$, $\Phi(\cdot)$ is the normal cumulative distribution, and $\phi(\cdot)$ is the normal density function.

When applied directly to our calibration problem, $\mu^\ast$ becomes critical in the expected outcome. 

In this paper, we propose using a simple Deep Learning Multi-Perceptron Network structure to capture a deterministic~\cite{polson_deep_2017} mean approximation of the surrogate function. Additionally, we intend to evaluate the influence of tuning the deep network dynamically as more candidate points are evaluated throughout the course of the calibration process. To denote the difference, a mean function trained only on the initial sample set prior to the calibration process will be denoted as $m(\theta)=DL_{Static}$; a mean function trained on the initial sample set prior to and then periodically tuned during the course of the calibration process will be denoted as $m(\theta)=DL_{Dynamic}$.

\section{Empirical Results}\label{Empirical}

Microscopic agent-based simulators for urban transportation systems include detailed models of traveler behavior and traffic flows; each agent becomes capable of reflecting a traveler's individual preferences, biases, and vehicle choices. In discretizing these characteristics, simulators can accurately represent land use variables, socio-demographic characteristics of the population, road infrastructure, and congestion patterns and have found wide use by researchers and urban planners for insights into usage behavior and forecasting impacts of regional development.

\begin{figure}[H]
	\centering
	\includegraphics[width=0.6\linewidth]{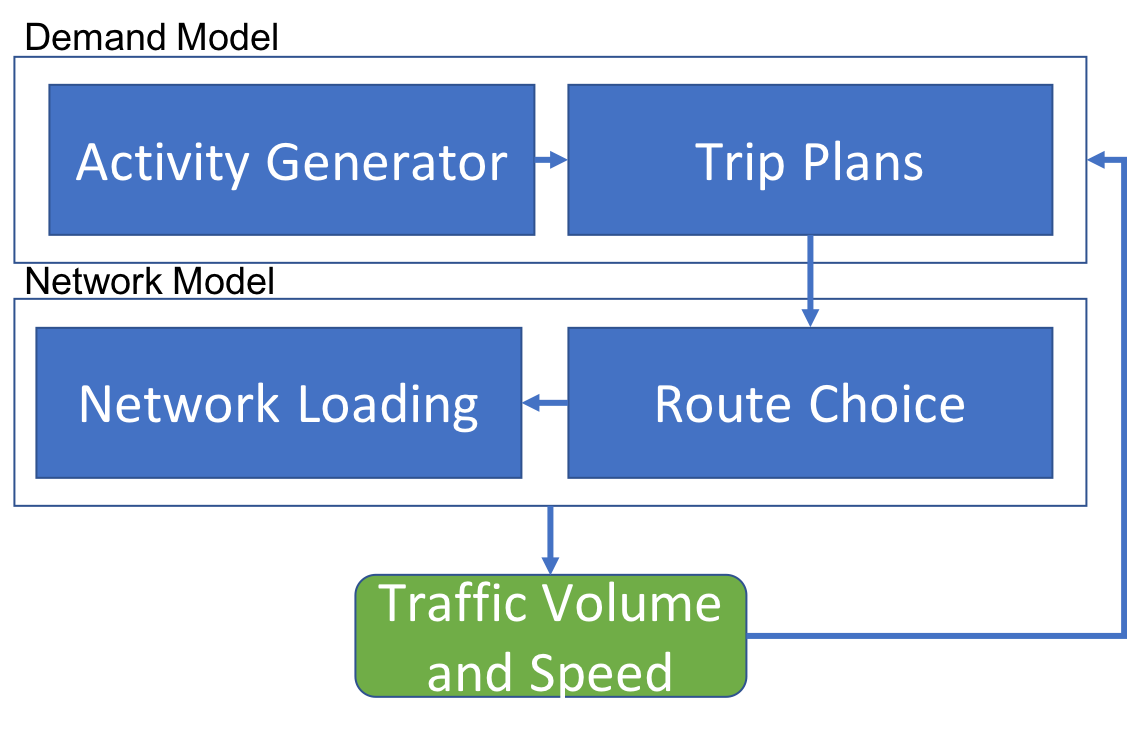}
	\caption{Integrated transportation simulator}
	\label{fig:polaris}
\end{figure}

We demonstrate our methodology on a problem of calibrating such an integrated transportation simulator known as POLARIS~\cite{auld_polaris:_2016}.  POLARIS includes two modules, the demand model and network model. Figure~\ref{fig:polaris} shows the overall architecture.

Hazard model is used for activity generation. The probability $h_{ij}(t)$ of an activity of type $j$ to occur for traveler $i$ at time $t$ is defined by the hazard function. Assuming Weibull baseline hazard functions, this probability is given by
\[
h_{ij}(t) = \gamma_{j} t^{\gamma-1}\exp(-\beta_j^Tx_i),
\]
where $\gamma_j$ is the shape parameter of the Weibull distribution, $x_i$ are observed attributes of the traveler and $\beta$ are the model parameters estimated from a survey. 

We ran the simulator several times to generate an initial sample set for use as training data to explore the relationship between the inputs and outputs; a hypercube sampling across the original dimensions subspace is utilized.

Trip planing model defines the order and timing within the simulation when the various attribute decisions are made. Then it, assigns attributes to each generated activity, such as location, start time, duration and mode. The planning order is modeled using the Ordered Probit model, which is an extension of the basic probit model for binary responses extended to ordinal responses $y \in \{1,\ldots K\}$. Under this model, the probability of $y$ being equal to $k$ is given by
\[
p(y = k) = p(l \le \alpha_k) - p(l\le \alpha_{k-1})
\]
%where $\alpha_1,\ldots\alpha_{K-1}$ is a set of threshold-values, and $l = \beta^Tx + \epsilon,~~~\epsilon\sim N(0,\Sigma)$.

At the time within the simulation when attributes are to be assigned for a given trip, a set of discrete choice models are used. For example, a multinational logit model~\cite{ben1985discrete} is used to choose destination for a given activity type. The model is derived from random utility maximization theory, which states that for each decision maker $n$ and location $i$, there is a utility $U_{in}$ associated with selecting location $i$. A generalized linear model for utility is assumed
\[
U_{in} = \beta^Tx_{ni} + \epsilon_{in},
\]
where $\epsilon_{in}$ is an unobservable random error, $\beta_i$ are the parameters estimated from survey data and $x_{ni}$ are observed attributes of traveler $n$ and location $i$, such as travel time to $i$, land use attributes of $i$. We assume $\epsilon_{in}$ follows Gimbel distribution, then the probability of traveler $n$ choosing destination $i$ is given by
\[
p_n(i) = \exp(V_{in})/\sum_{j \in C_n} \exp(V_{jn})
\]
here $V_{in} = \beta^Tx_{ni}$.  The choice set $C_n$ is formed using a space-time prism constraint~\cite{auld2011planning}.

Route choice is performed using Dijkstra's algorithm. Once the pre-trip choices on route, destination and departure time were assigned to agent, traffic flow simulator is used to model congestion level of road network.  Kinematic Wave theory of traffic flow~\cite{newell_simplified_1993} is used to simulate traffic flow on each road segment. This model has been recognized as an efficient and effective method for large-scale networks~\cite{lu_dynamic_2013} and dynamic traffic assignment formulations. In particular, the Lighthill-Whitham-Richards (LWR)  model ~\cite{lighthill_kinematic_1955,richards_shock_1956} along with discretization scheme proposed by Newell~\cite{newell_simplified_1993} is used. The LWR model is a macroscopic traffic flow model. It is a combination of a conservation law defined via a partial differential equation and a fundamental diagram. The fundamental diagram is a flow-density relation $q(x,t) = q(\rho(x,t))$.  The nonlinear first-order partial differential equation describes the aggregate behavior of drivers. The density $\rho(x,t)$ and flow $q(x,t)$, which are continuous scalar functions,  satisfy the equation 
\begin{equation*}\label{eqn:lwr1}
\frac{\partial\rho(x,t)}{\partial t} + \frac{\partial q(x,t)}{\partial x} = 0.
\end{equation*}
This equation is solved numerically by discretizing  time and space. 

In addition, intersection operations are simulated for signal controls, as well as stop and yield signs. The input for LWR model is so called-fundamental diagram that relates flow and density of traffic flow on a road segment. The key parameter of fundamental diagram is critical flow, which is also called road capacity. The capacity is measured in number of vehicles per hour per lane, and provides theoretical maximum of flow that can be accommodated by a road segment.

\subsection{Calibration Objective}

Due to limited computing availability, we treat only nine of the behavioral variables as unknown inputs,$\theta$, needing calibration to demonstrate the efficacy of the techniques outlined in this paper; the other behavioral variables not being calibrated are assumed to be known constants for the purpose of this example. 

The calibration framework's objective function is to minimize the mean discrepancy of the average simulated turn travel times across a 24 hour period between the calibrated inputs and the full set of true behavioral values. To quantify this discrepancy, the mean squared error function $L$ is used on the average turn travel times for every 5-minute interval during a 24-hour simulated period:
\begin{equation}
L(\theta,D)= \frac{1}{288}\sum_{i=1}^{288}(y - \psi(\theta_i))^2.
\end{equation}

\begin{figure}[H]
	\begin{tabular}{cc}
\includegraphics[width=0.45\linewidth]{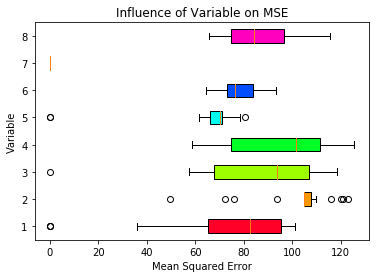} &
\includegraphics[width=0.55\linewidth]{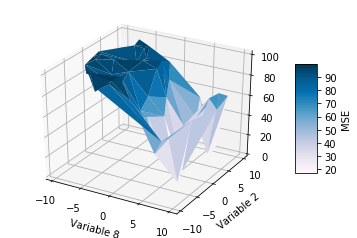} \\
		(a)Exploratory Analysis of Calibration Variables $\theta$   & (b) Interaction Influence of Variables 8 and 2 \\	
	\end{tabular}
	\caption{(a) A box plot indicating key statistics regarding each of the $9$ calibration variables. (b) A three-dimensional contour plot emphasizing the interaction term relationship between variables 8 and 2.}
	\label{fig:EDA}
\end{figure}

The possible range of values is assumed to be the same for every variable-- between -10 and 10. However, the influence among these calibration variables on the above Loss function, as shown in Figure \ref{fig:EDA}(a), is unique in both range and distribution. Most notably, the 8th variable appears to have no influence; however, interactions between the variables prove to also be influential. For example, variable 8 has a significant influence on the accuracy of the simulator when moved in combination with variable 2. See Figure \ref{fig:EDA}(b).

\subsection{Findings}

Python's threading module was utilized as the HPC program, which coordinates worker units to run the simulation code across multiple processors. The Bayesian optimization model exploration program, which utilizes the George~\cite{ambikasaran_fast_2014} Gaussian Process python package, evaluated points densely sampled using the Latin Hypercube method according to the Expected Improvement utility criterion in batch-sizes of 4. 

Finally, the calibration process was run using the following dimensional configurations: 

\begin{list}{$\bullet$}{}
		\item {\textbf{Original}} - the unadulterated $9$ dimensional subspace
		\item {\textbf{Static Active Subspace ($AS_S$)}} - the $n$ dimensional subspace resulting from the Active Subspace dimensionality reduction pre-processing
		\item {\textbf{Dynamic Active Subspace ($AS_D$)}} - the $n$ dimensional subspace resulting from the initial Active Subspace dimensionality reduction pre-processing and an updated Active Subspace dimensionality reduction performed every 2 iterations
		\item {\textbf{Static Deep Neural Network ($DL_S$)}} - the $nnnnn$ dimensional subspace resulting from the Deep Neural Network dimensionality reduction pre-processing
		\item {\textbf{Dynamic Deep Neural Network ($DL_D$)}} - the $nnnnn$ dimensional subspace resulting from the initial Deep Neural Network dimensionality reduction pre-processing and a tuned Deep Neural Network dimensionality reduction performed every 2 iterations
\end{list}

and GP configurations:
\begin{list}{$\bullet$}{}
	\item {\textbf{Zero Mean} ($m=0$)} A GP with a $0$ mean function and Mat\'ern covariance function over
	\item {\textbf{Static Deep Neural Network Mean} ($m=DL_S$)} A GP with a Deep Neural Network-derived mean function and Mat\'ern covariance function. The Deep Neural Network created for the mean function is trained once prior to the iterations.
	\item {\textbf{Dynamic Deep Neural Network Mean} ($m=DL_D$)} A GP with a dynamic Deep Neural Network-derived mean function and Mat\'ern covariance function. The Deep Neural Network created for the mean function is re-trained every second iteration
\end{list}

\begin{figure}[H]
	\begin{tabular}{cc}
\includegraphics[width=0.5\linewidth]{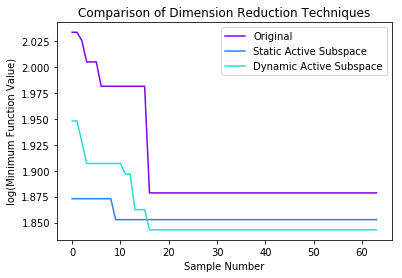} &
\includegraphics[width=0.5\linewidth]{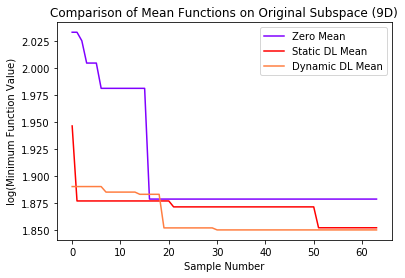} \\
		(a) Dimensional Reduction Influence & (b) Mean Function Influence \\	
	\end{tabular}
	\caption{(a) Comparison of dimensional reduction techniques on a zero-mean GP through Minimum Value Function in terms of Iterations. (b) Comparison of deep learner mean function through Minimum Value Function in terms of Iterations.}
	\label{fig:Compresults}
\end{figure}

As shown in Figure \ref{fig:Compresults}(a), the sole addition of dimension reduction pre-processing improved the calibration over the 60 iterations; likewise, Figure \ref{fig:Compresults}(b) demonstrates the positive influence provided by our improved mean function.

\begin{figure}[H]
	\begin{tabular}{cc}
\includegraphics[width=0.5\linewidth]{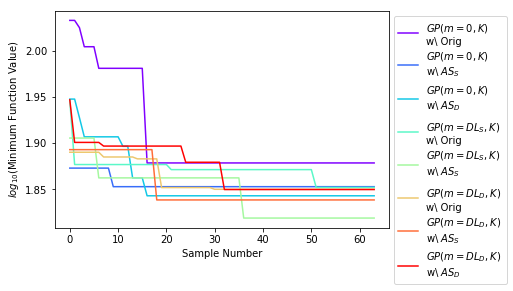} &
\includegraphics[width=0.5\linewidth]{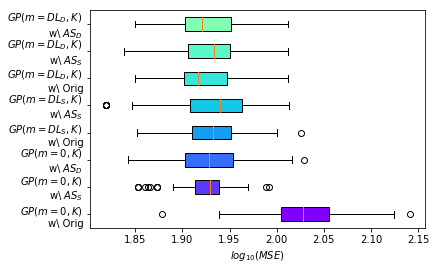}\\
		(a) Minimum Value Comparison & (b) Distribution of Samples by Technique\\	
	\end{tabular}
	\caption{(a) Comparison of Minimum Function Value across dimensional and GP configurations in terms of Iterations. (b) Summary Statistics regarding the distribution of the samples suggested by each technique.}
	\label{fig:results1}
\end{figure}

Of note, Figure \ref{fig:results1}(a) shows how the combination of reduced dimension and deep learner mean function quickly found a minimum function value while the original dimension method took several additional iterations before finding any significant reductions. The single addition of a static dimensional reduction or mean improvement significantly tightened the range in recommended sample results; however, the addition of any combination of dimension reduction and mean improvement proved beneficial among all sufficient statistics, as depicted in Figure \ref{fig:results1}(b).    

\begin{figure}[]
	\begin{tabular}{cc}
\includegraphics[width=0.5\linewidth]{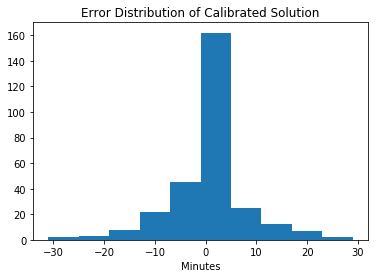} &
\includegraphics[width=0.5\linewidth]{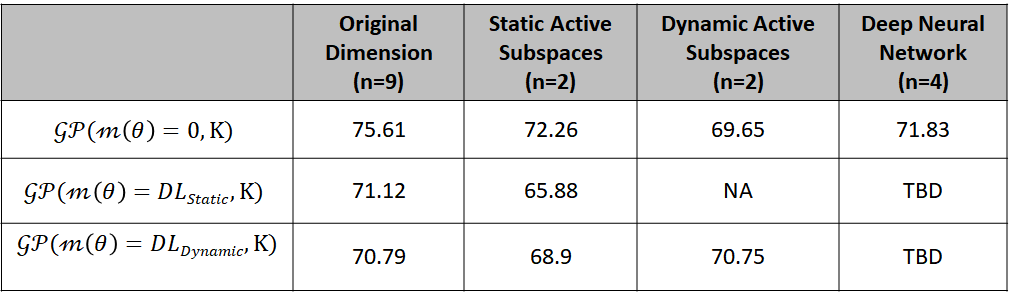} \\
		(a) Calibrated vs True   & (b) Minimum Loss Value Found by Calibration Technique  \\	
	\end{tabular}
	\caption{(a) Comparison of Calibrated and True Travel Time Outputs with Above Average Differences. (b) Optimal results of variable calibration using Bayesian optimization.}
	\label{fig:results2}
\end{figure}

Overall, the performance of the calibration framework provided a calibrated solution set which resulted in outputs, on average, within 6.4 minutes of the experiment's true output across the 24-hour period. With a standard deviation of 5 minutes, Figure \ref{fig:results1}(a) provides a visualization for those periods, roughly 15\% overall, which possessed greater than average variation from the true demand's output.

\section{Discussion}\label{Conclusion}
In this paper we provided a Bayesian optimization framework for complex transportation simulators. The goal of this paper is to show that Bayesain optimizaiton algorithms when combined with dimensionality reduction techniques is good default option for complex transportaiton simulators. Bayesian approach provide several advantages over other black-box optimization techniques, among those, are
\begin{enumerate}
	\item Distribution over functions provide a clear approach to the active learning problem (sequential design of computational experiment).
	\item The posterior distribution allows for sensetivity analysis and uncertainty quantification to be performed.  
\end{enumerate}

Although Bayesian optimization has been proposed before for simulation-based problems, our approach that combines it with dimensionality reduction techniques allow to solve real-life high-dimensional problems. Our parallel batching strategy and high performance computing framework allow for this approaches to be used by both researchers and transportation practitioners. 

%Our contributions are:
%\begin{itemize}
%	\item Formulation of Gaussian Process Bayesian methods for transportation optimization
%	\item Experiment Design algorithms for active conservation of restricted resource allocations
%	\item Dimensionality reduction techniques for complexity control
%	\item Integration of Deep Learning with Gaussian Process for improved mean function
%\end{itemize}
%
%We presented a framework for performing Bayesian calibration for transportation problems and introduced a Gaussian Process treatment in conjunction with active learning methods to maximize potential computational and time constraints. In addition, we have addressed multiple hindrances of the Gaussian Process through dimension complexity control and the integration of additional statistical methods through an improved mean function. Since our framework is input agnostic and can be used to calibrate other types of inputs, such as parameters of ABM models or land use variables, further work will include the application of this framework to additional implemented transportation simulators in higher dimensions.

\appendix
\section{Appendix}
\subsection{Calibration Algorithm}\label{Algorithm}

	\begin{algorithm}
	\caption{Calibration algorithm}\label{algorithm}
	\begin{algorithmic}[1]
		\Require{number of trials $t$, number of suggested candidates per trial $n$, variables to calibrate $\theta$, number of dimensions $d$, range of each dimension $[Lower_{1:d},Upper_{1:d}]$,dimension reduction type $DR$,mean type $mean$, and hyperparameters $\Psi$}
		\Procedure{Pre-Processing}{$DR,mean,\Psi,s_e,d,[Lower_{1:d},Upper_{1:d}]$}
		\State{\textbf{return} $s_e,d,[Lower_{1:d},Upper_{1:d}],\Psi$}
		\EndProcedure	
		\For{$l \gets 1$ to $t+1$}
			\If{$s_p \not= \{\}$}
				\For{$\forall i \ in \ s_p,\ simultaneously$}
					\State $y_i=Simulate(\theta_i)$
					\State $s_e.append([y_i,\theta_i])$
					\State $s_p.remove(\theta_i)$
				\EndFor
			\EndIf
			
			\If{$l<t+1$}
				\Procedure{Optimize}{$n,s_e,d,[Lower_{1:d},Upper_{1:d}]$}
					\State{$gp \gets \mathcal{GP} \mid \Psi,mean$}\Comment{Create a $d$ dimensional GP with hyperparameters $\Psi$}
					\If{$size(s_e)>10$}	\Comment{optimize hyperparameters after 10 samples}
						\State{$\min_\Psi -\log(gp_{likelihood})$}
					\EndIf
					\State{$pool_{1:2000} \gets LHS(d,[Lower_{1:d},Upper_{1:d}],2000)$}\Comment{$2000$ can be any large value}
					\For{$i \in 1 \ to \ n$}
						\State{$gp.compute(s_e)$}\Comment{Update posterior using evaluated samples}
						\State {$[\mu,K]= gp.predict(s_e,pool)$}\Comment{Determine predictive posterior}
						\State{$\theta_i^\ast=\argmax EI[U(pool)]$}\Comment{Next Sample maximizes the chosen acquisition}
						\State{$pool.remove(\theta_i^\ast)$}
						\State{$E[\theta_i^\ast] \gets rand() K[\theta_i^\ast]+\mu[\theta_i^\ast]$}
						\State{$s_e.append([\theta_i^\ast,E[\theta_i^\ast]])$}
					\EndFor
					\State{\textbf{return} $\theta^\ast_{1:n}$}
				\EndProcedure	
				\State $s_p.append(\theta_{1:n}^\ast)$
			\EndIf
		\EndFor
		\State{\textbf{return} $\theta^+ = \argmin_{\theta_i \in s_e} y_i$}
	\end{algorithmic}
\end{algorithm}

\begin{algorithm}
	\caption{Pre-processing algorithm}\label{algorithm2}
	\begin{algorithmic}[1]
		\Require{number of dimensions $d$, range of each dimension $[Lower_{1:d},Upper_{1:d}]$,dimension reduction type $DR$,mean type $mean$, hyperparameters $\Psi$}
		\Procedure{Pre-Processing}{$DR,mean,\Psi,s_e,d,[Lower_{1:d},Upper_{1:d}]$}
			\State{$set_{1:q} \gets LHS(d,[Lower_{1:d},Upper_{1:d}],q)$, $2d \le q \le 10d$}\Comment{Latin Hypercube Sample(LHS)}
			\For{$i \in 1 \ to \ q$}
				\State $y_i=Simulate(set_i)$
				\State $s_e.append([y_i,set_i])$
			\EndFor
			\If{$DR=="Active \ Subspaces"$}
				\State{$AS=FindActiveSubspaces(s_e,[Lower_{1:d},Upper_{1:d}])$}
				\State{$d \gets AS(d)$}\Comment{update $d$ to new reduced dimension}
				\State{$[Lower_{1:d},Upper_{1:d}] \gets AS([Lower_{1:d},Upper_{1:d}])$}\Comment{update bounds to new reduced dimension bounds}
				\State{$s_e \gets AS(s_e)$}\Comment{Convert evaluated samples to new reduced dimensions}
			\ElsIf{$DR=="Neural \ Network"$}
				\State{$NN=FindNeuralNetwork(s_e)$}
				\State{$d \gets NN(d)$}\Comment{update $d$ to new reduced dimension}
				\State{$[Lower_{1:d},Upper_{1:d}] \gets NN([Lower_{1:d},Upper_{1:d}])$}\Comment{update bounds to new reduced dimension bounds}
				\State{$s_e \gets NN(s_e)$}\Comment{Convert evaluated samples to new reduced dimensions}
			\EndIf
			\If{$mean=="DL"$}
				\State{$\Psi.append(m(\theta)=TrainNeuralNetwork(s_e))$}
			\Else
				\State{$\Psi.append(m(\theta)=0)$}
			\EndIf
			\State{\textbf{return} $s_e,d,[Lower_{1:d},Upper_{1:d}],\Psi$}
		\EndProcedure	
	\end{algorithmic}
\end{algorithm}

\bibliographystyle{plain}
\bibliography{../../ref2}

\end{document}